\documentstyle[epsf,11pt]{article}

\begin{document}

\renewcommand{\thefootnote}{\fnsymbol{footnote}}
\begin{flushright}
  {\bf TTP97-38\footnote{The complete paper, including
      figures, is also 
      available via anonymous ftp at
      ftp://ttpux2.physik.uni-karlsruhe.de/, or via www at\\
      http://www-ttp.physik.uni-karlsruhe.de/cgi-bin/preprints/}}\\
 {\bf hep-ph/9710287}\\
October 1997
\end{flushright}
\vspace*{15mm}
\begin{center}
\boldmath
{\bf \Large Probing the helicity structure of $b\to s\gamma$ 
       in $\Lambda_b \to \Lambda \gamma$ }\\[1cm]
\unboldmath
Presented at XXIst School of Theoretical Physics, Ustron 97\\[5mm]
{\bf\large Thomas Mannel and Stefan Recksiegel}\\[5mm]
{\em Institut f\"ur Theoretische Teilchenphysik, 
         Universit\"at Karlsruhe,\\ D-76128 Karlsruhe, Germany\\[2mm]}
\end{center}
\thispagestyle{empty}

\begin{abstract}
We investigate the rare decay $\Lambda_b \to \Lambda \gamma$ which receives
both short and long distance contributions. We estimate the long distance
contributions and find them very small. The form factors are obtained
from $\Lambda_c \to \Lambda \ell \bar{\nu}_\ell$ using heavy quark symmetry
and a pole model. The short distance piece opens a window to new physics
and we discuss the sensitivity of  $\Lambda_b \to \Lambda \gamma$
to such effects.

\end{abstract}

\makeatletter
\def\fmslash{\@ifnextchar[{\fmsl@sh}{\fmsl@sh[0mu]}}
\def\fmsl@sh[#1]#2{%
  \mathchoice
    {\@fmsl@sh\displaystyle{#1}{#2}}%
    {\@fmsl@sh\textstyle{#1}{#2}}%
    {\@fmsl@sh\scriptstyle{#1}{#2}}%
    {\@fmsl@sh\scriptscriptstyle{#1}{#2}}}
\def\@fmsl@sh#1#2#3{\m@th\ooalign{$\hfil#1\mkern#2/\hfil$\crcr$#1#3$}}
\makeatother
\def\g5{\gamma_5}
\def\ml{m_\Lambda}
\def\mb{m_{\Lambda_b}}
\def\beq{\begin{equation}}
\def\eeq{\end{equation}}
\def\beqa{\begin{eqnarray}}
\def\eeqa{\end{eqnarray}}
\def\ps{{\rm{\bf\hat{p}.S}_\Lambda}}

\section{Introduction}
\thispagestyle{empty}
Flavour Changing Neutral Current (FCNC) processes have attracted
renewed attention since the recent CLEO measurement of the FCNC decays
of the type $b \to s \gamma$. Since in the Standard Model (SM) these
processes are forbidden at the tree level and hence are strongly suppressed
by the GIM mechanism, they
offer a unique possibility to test the CKM sector of the SM.

Based on the decays of $B$ mesons it will not be possible to analyse
the helicity structure of the effective hamiltonian mediating the
decay $b \to s \gamma$, since the information on the handedness of the
quarks is lost in the hadronization process.
The only chance to access the helicity of the quarks is to consider
the decay of baryons.

Heavy--to--light transitions between ground state baryons are
in large parts of the phase space restricted by heavy quark symmetries.
For the baryonic transition 
$\Lambda_Q \to $ light spin-1/2 baryon the number of independent
form factors is restricted to only two. However, it is expected
that heavy quark symmetries work best for such a kinematic configuration
where the outgoing light hadron is almost at rest,
corresponding to the point of maximum momentum transfer.
For the decay under consideration, $b\to s \gamma$, we are
at the opposite side of phase space where $q^2=0$.
We shall still use the relations implied by heavy quark
symmetry although there is no {\it a priori} reason to expect that
they hold at $q^2=0$. This is motivated from models based on the
diquark picture which phenomenologically work quite well.

Another theoretical difficulty with the process suggested is possible
long distance contributions which will dilute the effects of a
non--SM--\-contribution to the short distance effective hamiltonian 
for $b\to s \gamma$.
Such contributions can only be estimated in terms of models but they
turn out to be small.

\section{The decay amplitude for $\Lambda_b \to \Lambda \gamma$}
The effective hamiltonian for this specific decay consists in the
SM of only one operator, in which (up to tiny effects from the
strange quark mass) the $b$--quark is right--handed
and the $s$--quark is left--handed. Generalizing this we introduce
couplings constants $g_V$ and $g_A$ such that the effective hamiltonian takes
the form:
\beq
H_{\rm eff}={4G_F\over \sqrt{2}}V_{ts}^*V_{tb} C_7 {\cal O}_7, \quad
{\cal O}_7 = \frac{e}{32 \pi^2} m_b \bar{s} \sigma_{\mu \nu}
             (g_V-g_A\gamma_5) b F^{\mu \nu}
\eeq
In the Standard Model $g_V=1+{m_s/ m_b}, g_A=-1+{m_s/ m_b}$ and $C_7 =0.325$
from a leading log QCD calculation.
We shall use $C_7 g_V$ and 
$C_7 g_A$ as parameters describing strength and helicity structure of the
short distance process.

It remains to calculate matrix elements of the operators. In the present case
we shall make use of the fact that the $b$ quark is heavy, while the $s$ 
quark is taken to be a light quark. In general, heavy quark symmetries  
restrict the number of possible form factors quite significantly; in the 
present case there are only two, parametrized by
\cite{baryonsinhqet}
\beq \label{heavytolightdecay}
\left< \Lambda(p,s)|\bar s \Gamma b | \Lambda_b(v,s') \right>
= \bar u_\Lambda (p,s) \{ F_1(p.v)+\fmslash v F_2(p.v)\} \Gamma u_{\Lambda_b}
(v, s') \eeq
$\Gamma$ is an arbitrary Dirac matrix, such that any 
transition between a $\Lambda_Q$ ($Q = b,c$) and a light spin $1/2$ baryon 
is given by the same form factors. As mentioned in the introduction,
(\ref{heavytolightdecay}) is expected to work best at $q^2_{\rm max}$,
but we shall use this relation also at $q^2=0$, which is in the worst
case a model assumption.

The quantity of interest is the decay rate of unpolarized $\Lambda_b$ 
baryons into $\Lambda$ baryons with a definite spin directions $s$.
This rate can be written in terms 
of the polarization variables as defined in \cite{pdg} 
\beq
\Gamma = \Gamma_0 \cdot [ 1 + \alpha' \ps]
\eeq
where ${\rm\bf\hat{p}}$ is the momentum vector of the $\Lambda$
and ${\rm{\bf S}_\Lambda}$ is its spin vector. 
Here $\Gamma_0$ depends on the sum $g_V^2 + g_A^2$ only and the other
dependence on $g_V$ and $g_A$ is contained in
\beqa
\alpha'
&=& 0.378 \cdot {2g_V g_A \over g_V^2 + g_A^2}. 
\label{alphadef} \eeqa
The polarization variable depends only on the ratio of the
form factors for which we have assumed a constant value and which
has been extracted by CLEO to be
\cite{cleo} $R={F_2/ F_1}= -0.25 \pm 0.14 \pm 0.08$.
Furthermore we have used
he ratio of the baryon masses $x=0.20$.

In addition to the short distance part there are also long distance
contributions to $\Lambda_b \to \Lambda \gamma$: One type of
long distance contributions involves a virtual $J/\Psi$ that
subsequently decays into the final state photon, this is a
vector meson dominance like contribution.
The other type of short distance contribution arises from a
$W$--boson being exchanged between two of the internal quark
lines, the photon is radiated off any of the internal quark lines.

We have estimated both these contributions employing a
vector dominance type model in combination with factorization
and a simple diquark picture for the internal $W$--exchange.
Both contributions are small;
the former can be expressed as a correction to the short distance
couplings $g_V$ and $g_A$ (magnitude indicated by the width of
the lines in Fig.\ref{gagvfig}) and the latter one is negligible.

\section{The decay rate and polarization asymmetry for
         $\Lambda_b \to \Lambda \gamma$}
We extract the relevant form factors from the decay 
$\Lambda_c \to \Lambda \ell \bar{\nu}_\ell$, which is according to 
(\ref{heavytolightdecay}) given by the same two form factors as
 $\Lambda_b \to \Lambda \gamma$. To do so we must make a model
for the $q^2$--dependence of the form factors.
Starting from a simple pole model, 
\begin{equation} 
F_{1/2}(q^2)=F^{\rm max}_{1/2} \cdot \left({M_{1/2}^2-m_Q^2
\over M_{1/2}^2-q^2}\right)^n, \quad n=1,2\, , \label{dipol} 
\end{equation}
where $M_{1/2}$ is the mass of the nearest resonance with the correct 
quantum numbers for $F_1$ and $F_2$, we use the relation between
$q^2$ and $v.p$ as well as heavy quark symmetry to obtain

\begin{equation} \label{poleforms}
F_{1/2}(p.v)=N_{1/2} \cdot \left({\Lambda
\over \Lambda+p.v}\right)^n, \quad n=1,2\, , 
\end{equation}
where $\Lambda = 200 MeV$ is found from the masses of the respective
resonances for both form factors. 
In (\ref{dipol}) and (\ref{poleforms}) $n=1$ and $n=2$ denote
monopole and dipole phase space depencence, respectively. We
shall give our final results for both these models.
For simplicity we have assumed
the same $v.q$--dependence for both of the form factors which
again allows us to use the phase space averaged measurement of
$\left<F_2/F_1\right> = -0.25 \pm 0.14 \pm 0.08  \cite{cleo}$
for the ratio of $F^{\rm max}_{1}$ and $F^{\rm max}_{2}$.

Using the data for rate for 
$\Lambda_c \to \Lambda \ell \bar{\nu}_\ell$ \cite{pdg}
we obtain an expression for the total decay rate of
$\Lambda_b \to \Lambda \gamma$
in which only the depencence on the fifth power of the mass of
the $b$ quark introduces a sizable uncertainty. To eliminate this
uncertainty we compare the decay $\Lambda_b \to \Lambda \gamma$ to
the semileptonic decay of the $B$ meson which exhibits the same
$m_b^5$ dependence.
Using the lifetimes and the inclusive semileptonic branching fraction
$BR(B \rightarrow X_c l\nu)=(10.3 \pm 1.0) \%$
from \cite{pdg} we obtain
\beq
BR(\Lambda_b \rightarrow \Lambda \gamma)= 
    (1- 4.5) \cdot 10^{-5},
\eeq
where the lower and upper value correspond to the dipole and monopole
$q^2$ evolution of the form factors, respectively. 

For values of $g_A$ and $g_V$ different from the standard model ones
one has to multiply the above equation by the factor $(g_A^2 + g_V^2)/2$, 
neglecting the strange quark mass and the long distance contributions.

In the following we shall use the measurement of the inclusive rate 
$B \to X_s \gamma$ to fix $(g_V C_7)^2+ (g_A C_7)^2 $; 
in the $C_7 g_V$--$C_7 g_A$--plane a measurement of any total rate (meaning 
any of the processes $B \to X_s \gamma$, $B \to K^* \gamma$ or
$\Lambda_b \to \Lambda \gamma$) will correspond to a circle, since all 
these decay rates are proportional to $(g_V C_7)^2+ (g_A C_7)^2 $. 

If in future experiments a measurement of $\alpha'$ in 
$\Lambda_b \to \Lambda \gamma$ is performed one may use the relation 
between $\alpha'$ and $(C_7 g_A)/(C_7 g_V)$. Including our  
estimate of the long distance contributions, one obtains 
\begin{equation}
C_7 g_A = (C_7 g_V + C_7 \rho) 
          \left(\frac{0.353}{\alpha'} \pm 
          \sqrt{\left(\frac{0.353}{\alpha'}\right)^2 - 1} \right) 
                   + C_7 \rho
\end{equation}
which corresponds to straight lines in the $C_7 g_A$--$C_7 g_V$ plane. 

\begin{figure}[ht]
\vspace{-.5cm}
 \begin{center}
  \leavevmode
  \epsfxsize=8cm
  \epsffile[70 160 540 630]{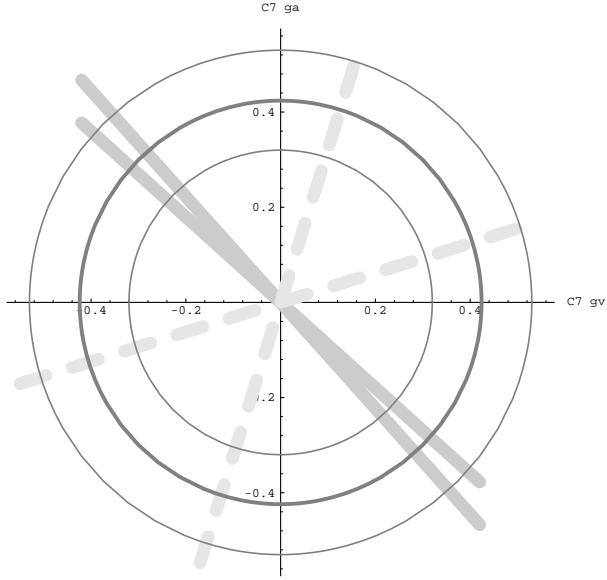}
 \end{center}
\vspace*{-.5cm}
\caption{The $C_7 g_A$--$C_7 g_V$ plane \label{gagvfig}}
\end{figure}
In fig.\ref{gagvfig} we plot the $C_7 g_A$--$C_7 g_V$ plane. 
The central circle corresponds to the central value of the measurement 
of $B \to X_s \gamma$ and the thin circles indicate the experimental 
uncertainty. Assuming the standard model values for $C_7 g_A$ and 
$C_7 g_V$ we have also plotted the corresponding lines. The width of 
these lines is given by our estimate for the long distance 
contributions; we have taken this estimate as an additional 
uncertainty meaning that the width of these lines is $2\rho C_7$. 

A measurement of $\alpha'$ yields in general two lines which have 
in total four intersections with the circle. The two intersections
of each line correspond to a sign interchange $g_A \to -g_A$ 
and at the same time $g_V \to -g_V$ which is an unobservable phase. 
The remaining ambiguity corresponds to the interchange 
$g_A \to g_V$ and $g_V \to g_A$, since the polarization variable 
and the total rate are symmetric functions of $g_A$ and $g_V$. 
Graphically this means that the two lines for a measured value of 
$\alpha '$ are mirror images of each other with respect to the lines 
$g_A = |g_V|$. In order to resolve this ambiguity additional 
measurements would be necessary. 

In the SM $|g_A| \approx |g_V|$ ($\alpha ' = -0.351$)
up to corrections from the non-vanishing $s$ quark mass,
therefore the two solid lines almost coincide for the SM value of
$\alpha'$. For illustrations we also have plotted two dashed lines for a hypothetical 
measurement of $\alpha ' =0.2$.

With a branching ratio for $\Lambda_b \to \Lambda \gamma$ of the 
order $10^{-5}$ one needs $10^8$ $b$ quarks to have about one hundred 
events, without applying cuts for efficiencies. Clearly this will 
be feasible at dedicated $b$ physics experiments at colliders 
such as the one at Tevatron or LHC, and possibly also at fixed 
target experiments like HERA-B.

\end{document}